# Multi-Modal Information Fusion of Acoustic and Linguistic Data for Decoding Dairy Cow Vocalizations in Animal Welfare Assessment

Bubacarr Jobarteh[1], Madalina Mincu[2], Gavojdian Dinu[2], Suresh Neethirajan[1,3,*]

[1]Faculty of Computer Science, 6050 University Avenue, Dalhousie University, Halifax, Canada
[2]Cattle Production Systems Laboratory, Research and Development Institute for Bovine, Balotesti, Romania
[3]Faculty of Agriculture, Agricultural Campus, PO Box 550, Dalhousie University, Truro, NS, Canada B2N 5E3

*Corresponding Author. E-mail address: sneethir@gmail.com (Suresh Neethirajan)

**Abstract**
Understanding animal vocalizations through multi-source data fusion is crucial for assessing emotional states and enhancing animal welfare in precision livestock farming. This study aims to decode dairy cow contact calls by employing multi-modal data fusion techniques, integrating transcription, semantic analysis, contextual and emotional assessment, and acoustic feature extraction. We utilized the Natural Language Processing-based WHISPER model to transcribe audio recordings of cow vocalizations into written form. By fusing multiple acoustic features—frequency, duration, and intensity—with transcribed textual data, we developed a comprehensive representation of cow vocalizations. Utilizing data fusion within a custom-developed ontology, we categorized vocalizations into high-frequency calls (HFC) associated with distress or arousal, and low-frequency calls (LFC) linked to contentment or calmness. Analyzing the fused multi-dimensional data, we identified anxiety-related features indicative of emotional distress, including specific frequency measurements and sound spectrum results. Assessing the sentiment and acoustic features of vocalizations from 20 individual cows allowed us to determine differences in calling patterns and emotional states. Employing advanced machine learning algorithms—Random Forest, Support Vector Machine (SVM), and Recurrent Neural Networks (RNN)—we effectively processed and fused multi-source data to classify cow vocalizations, achieving accuracies of 97.25% (Random Forest), 98.35% (SVM), and 88.00% (RNN). The F1-scores for distress/arousal were 0.98 (Random Forest) and 0.99 (SVM). These models were optimized to handle computational demands and data quality challenges inherent in practical farm environments. Our findings demonstrate the effectiveness of multi-source data fusion and intelligent processing techniques in animal welfare monitoring. By interpreting cow vocalizations through information fusion, we can enhance welfare surveillance in precision livestock farming, making it more ethical and efficient. This study represents a significant advancement in animal welfare assessment, highlighting the role of innovative fusion technologies in understanding and improving the emotional well-being of dairy cows.

## 1. Introduction

Vocalizations in dairy cows are multifaceted expressions influenced by emotional states, environmental conditions, and management practices. They serve as crucial indicators of both positive and negative welfare states, reflecting a spectrum of emotions from distress and fear to contentment and social bonding. There is ample evidence that a cow's emotional state significantly influences its vocalizations. Specifically, vocalization rates and their acoustic structures—frequency, amplitude, and duration—vary predictably in response to emotional arousal; as arousal levels increase, sounds are produced more rapidly, with greater amplitude and at higher frequencies (Briefer, 2012). This consistent variation underscores the potential of vocalization analysis as a non-invasive tool for assessing animal welfare.

In the context of information fusion, integrating multiple acoustic features and environmental data can enhance the interpretation of these vocalizations. Low-frequency calls (LFCs) are typically associated with calm or social situations, emitted during close contact with the mouth closed or half-closed, facilitating social bonding within the herd. In contrast, high-frequency calls (HFCs) are used for long-distance communication, produced with the mouth open, and are indicative of negative emotional states (Briefer, 2012). These vocalizations manifest various feelings, highlighting the complexity of bovine emotional expression. By fusing data from different acoustic parameters and contextual factors, we can develop more accurate models for emotional state detection.

Environmental factors play a significant role in influencing cow vocalizations. Cows on pasture may vocalize differently than those housed indoors due to variations in acoustic properties, social interactions, and exposure to natural elements. Pasture environments facilitate more natural behaviors, potentially affecting the frequency and type of vocalizations (Mac et al., 2022). For instance, cows in open fields might produce more LFCs due to increased opportunities for social interactions and reduced environmental stress. Conversely, indoor systems present unique acoustic properties that impact cow communication, including alterations in sound-producing behavior related to extreme temperatures, storms, or confined spaces. Feeding times and social structures that vary between systems also influence vocalizations related to hunger or social interactions. Understanding these differences is vital for accurate interpretation of vocal behavior and tailoring welfare assessments to specific farming contexts (Mac et al., 2022). By integrating environmental sensors and contextual data, we can enhance the fusion of information to better assess welfare states.

Research consistently shows that vocal parameters vary significantly during positive and negative experiences (Silva et al., 2008). In situations of stress or discomfort, cows tend to produce vocalizations with higher frequencies, greater amplitudes, and longer durations. These acoustic features serve as reliable indicators of their emotional and physiological states. Recent advancements in the field of animal linguistics emphasize the importance of applying linguistic frameworks to animal communication systems (Berthet et al., 2023). This aligns with our approach of utilizing information fusion techniques, combining natural language processing (NLP) with acoustic feature extraction to analyze cow vocalizations, allowing for a more nuanced understanding of their emotional expressions.

Olczak et al. (2023) highlight the significance of sound in livestock farming, underscoring the value of acoustic analysis for a deeper understanding of cow behavior and welfare. Their work situates our findings within a broader context of farm management practices, emphasizing the potential for integrating acoustic monitoring into routine welfare assessments. Furthermore, Whitham and Miller (2024) demonstrate the potential of using vocalizations to gain insights into the affective states of non-human mammals, directly supporting our methodology of linking fused acoustic features to emotional states in cows. Collectively, these studies underscore the importance of our research in advancing animal welfare assessment and precision livestock farming through multi-source information fusion. They also suggest avenues for future research, such as comparing cow vocalizations with those of other livestock or examining how farm soundscapes affect vocalization patterns.

Over recent decades, there has been a significant shift in animal welfare monitoring practices. Traditional methods such as visual inspections, physiological assessments, and behavioral testing have been instrumental in improving animal care by providing fundamental insights into welfare (Neethirajan, 2024). However, these methods can be subjective and dependent on the observer's expertise, potentially altering the animals' natural behavior due to human presence. For example, the presence of human observers might suppress natural vocalizations or induce stress, thereby masking signs of pain or discomfort (Neethirajan, 2020). Vocalizations in dairy cows are significant components of their behavioral expressions, serving as valuable indicators of their emotional and physiological states. Changes in vocal patterns, particularly under stress or distress, can reveal critical insights into their well-being.

Acoustic features of animal vocalizations—such as frequency, duration, and amplitude—are critical indicators related to communication and emotions (Brudzynski, 2013; Moshou et al., 2001). These features enable the detection of changes in emotional states, contributing to welfare assessment and behavioral monitoring (McGrath et al., 2017; Meen et al., 2015). Parameters related to acoustic attributes are essential for developing automated welfare-monitoring systems and improved management strategies in animals (McLoughlin et al., 2019). By analyzing the frequency and temporal patterns of calls, along with their amplitude, and fusing this data with contextual information, scientists can determine the emotional states of animals, distinguishing between satisfaction, fear, stress, and social cohesion (Neethirajan, 2024). For instance, increased vocalization frequency and amplitude might indicate heightened stress levels, while more regular, lower-frequency calls could signify a state of contentment.

Acoustic analysis is highly valuable due to its non-invasive and objective nature in understanding the affective states of animals, playing a pivotal role in the continuous improvement of animal welfare. Implementing innovative technologies in acoustic analysis is crucial for developing effective and humane animal care techniques (Neethirajan, 2024). The integration of machine learning, NLP techniques, and multi-source data fusion offers new opportunities to automate and enhance the analysis of animal vocalizations. Advanced algorithms can facilitate the identification and classification of vocal patterns associated with specific emotional states by fusing acoustic features with semantic information, improving the accuracy and efficiency of welfare assessments.

The objective of this study is to investigate cow contact calls through multi-modal information fusion involving transcription and semantic analysis, contextual and emotional assessment, and acoustic feature extraction, comparing these elements across individual subjects. By employing NLP-based models and fusing acoustic data, we aim to develop a comprehensive framework for decoding bovine vocalizations, ultimately enhancing animal welfare monitoring in precision livestock farming through advanced information fusion techniques.

Table 1. Parameters of Vocalization Extracted from the Cow Sounds for Analysis

| Parameter | Definition | References |
|---|---|---|
| **Duration** | The duration of a vocalisation. | (Briefer et al. 2015a; Maigrot et al. 2018; Briefer et al. 2019; Friel et al. 2019) |
| **Vocalisation rate** | The number of vocalisations in a certain time frame. | (Stěhulová et al. 2008; Imfeld-Mueller et al. 2011; Briefer et al. 2019) |
| **F0** | The fundamental frequency and its contour (e.g. min, mean, max and range). | (Briefer et al. 2015a, b; Leliveld et al. 2016; Riley et al. 2015) |
| **FMextent** | The variation between two peaks of each F0 modulation in Hz. | (Briefer et al. 2015b) |
| **Bandwidth** | The difference between the highest and lowest observed frequency (Hz). | (Leliveld et al. 2016) |
| **Amplitude** | Level of energy in the vocalisation, the intensity of a vocalisation (decibel). | |
| **AMextent** | The mean-to-mean peak variation of each amplitude modulation (decibel). | (Briefer et al., 2015a; Maigrot et al., 2017) |
| **AMrate** | The number of amplitude modulations in a certain time frame. | (Maigrot et al. 2018) |
| **AMVar** | The cumulative variation in amplitude divided by the duration of a vocalisation (dB/s). | (Briefer et al. 2015a) |
| **Q25 %** | The frequency below which 25 percent of the energy is contained (Hz). | (Briefer et al. 2015a; Leliveld et al. 2016; Maigrot et al., 2017, 2018) |
| **Q50 %** | The frequency below which 50 percent of the energy is contained (Hz). | (Briefer et al. 2015a; Leliveld et al. 2016; Maigrot et al. 2018) |
| **Q75 %** | The frequency below which 75 percent of the energy is contained (Hz). | (Briefer et al. 2015a; Maigrot et al. 2017, 2018) |

| Parameter | Definition | References |
|---|---|---|
| **Formants** | Frequencies that correspond to the resonances of the vocal tract. | |
| **F1mean, F2mean, F3mean, F4mean** | The mean frequency of each formant (Hz). | (Maigrot et al. 2018) |
| **F1, F2, F3 and F4 range** | The frequency range of each formant, thus the difference between the maximum and minimum frequency of that formant (Hz). | (Maigrot et al. 2018) |
| **Fpeak** | The frequency of peak amplitude. | |

The vocalization recordings were analyzed using the ((Praat DSP package (v.6.0.31), Boersma and Praat, 2022) along with custom-built scripts previously developed by Briefer et al., 2015; Briefer et al., 2022; Reby and McComb, 2003; Gabriel, 2004 and Briefer et al., 2019, to automatically extract the acoustic features for each vocalization. The studied vocal parameters, along with their definitions, are presented above (Table 1).

## 2. Background and Related Work

Advancements in information fusion techniques have opened new avenues in bioacoustic signal processing and animal welfare assessment. OpenAI's Whisper model, a state-of-the-art speech recognition system based on transformer architecture, exemplifies such progress by demonstrating exceptional results in transcribing human speech (Radford et al., 2022). The model's architecture comprises an encoder and a decoder, where the encoder extracts meaningful representations from raw audio inputs through convolutional layers, self-attention mechanisms, and residual connections, efficiently converting audio signals into high-level contextual features.

Evidence indicates that Whisper excels even with noisy and unstructured data. It outperforms traditional models in processing noisy bioacoustic data and voice activity detection across datasets involving humans and animals, achieving higher accuracy under poor or mixed audio quality conditions (Hahnloser & Gu, 2023). The BioDenoising project further demonstrated that models like WhisperSeg, derived from Whisper, exhibit excellent performance in denoising animal vocalizations without access to clean data, confirming Whisper's effectiveness in noisy environments (Miron et al., 2024). Researchers have fine-tuned the Whisper model for specific bioacoustic tasks, such as bird call classification, significantly improving classification scores and broadening its applications in analyzing animal vocalizations (Sheikh et al., 2024).

Understanding the emotional associations of cow vocalizations is crucial for interpreting their behavioral states, and integrating multi-source data enhances this interpretation. High-frequency calls (HFCs) are linked to distress or arousal; these louder, longer, and higher-pitched sounds are associated with fear, isolation, or physical discomfort. Bach et al. (2022) observed that farmers reported more frequent vocalizations around dry-off periods, correlating with other signs of distress like reduced lying time, suggesting that vocalizations are reliable indicators of discomfort or stress. Similarly, Green et al. (2021) found that during distress, cows produced longer call sequences with greater cumulative vocalization durations, often consisting of repeated calls indicative of sustained distress.

Conversely, low-frequency calls (LFCs) correlate with contentment or calmness. Cows emit soft, low-pitched sounds when relaxed, particularly near their calves in the first few weeks postpartum. These calls are relatively quiet and are made with the mouth closed or partially open (De la Torre & McElligott, 2017). Essentially, HFCs serve the biological role of long-distance communication and indicate distress, whereas LFCs are primarily used for close-contact communication (De la Torre et al., 2014). By fusing acoustic features with contextual information, we can enhance the detection and interpretation of these emotional states.

In the context of information fusion, integrating acoustic features with advanced machine learning models offers a promising approach to decoding bovine vocalizations. Detailed information on experimental design, animal handling, data collection, audio equipment, and bioacoustics software used in this study has been previously published (Gavojdian et al., 2024).

### 3. Methodology and Data

Our study employs multi-modal information fusion techniques to analyze cow vocalizations, integrating acoustic feature extraction with advanced machine learning models to classify emotional states. We utilized a dataset of 1,144 vocalizations from 20 cows isolated for four hours post-milking to investigate frustration in cow vocalizations. By fusing features such as frequency, formants, and duration extracted from both raw four-hour recordings and segmented vocal events, we aimed to detect frustration buildup by examining shifts in these features over time.

3.1 Data Collection

We recorded 1,144 vocalizations from 20 cows under controlled isolation conditions, isolating each cow from the herd for four hours immediately after milking. Continuous vocalizations were recorded and segmented into individual events using Praat software, which extracted 23 acoustic parameters per vocalization, including fundamental frequency (F0), formant frequencies, duration, and harmonicity. The primary goal was to assess whether frustration modifies vocal parameters over time. By segmenting the four-hour recordings, we compared acoustic features between early and late isolation periods, identifying systematic shifts in vocal patterns. All raw recordings were analyzed, and trimmed sounds captured under negative emotional states due to isolation were included to maintain consistency throughout the dataset.

3.2 Data Preparation

Data preparation involved installing necessary software and libraries for audio processing and data manipulation, utilizing Python libraries such as NumPy, Pandas, and Scikit-learn. We performed audio normalization and noise reduction, segmenting vocalizations for further analysis. The OpenAI Whisper model transcribed cow vocalizations into text, and feature extraction was conducted using the Librosa library, focusing on parameters like frequency, duration, and loudness. By fusing these acoustic features with the transcribed data, we created a comprehensive representation of each vocalization. Vocalizations were categorized into high-frequency calls (HFC), labeled "Distress/Arousal," and low-frequency calls (LFC), labeled "Contentment/Calm," based on literature trends.

3.3 Feature Representation and Machine Learning Models

Feature representation was critical for classifying cow vocalizations based on fused acoustic properties. Using Librosa, we extracted key features such as frequency, duration, loudness, and formants from the dataset. These features were selected for their relevance in distinguishing between LFC and HFC, corresponding to emotional states like "Contentment/Calm" and "Distress/Arousal," respectively. We employed machine learning algorithms—including Random Forest, Support Vector Machine (SVM), and Recurrent Neural Network (RNN) models—to process and fuse the multi-dimensional data effectively.

Features were normalized and compiled into a feature matrix serving as input for the models. Each call was represented by a set of fused acoustic feature vectors, enabling classification into distinct emotional categories. The models were trained and validated using labeled data to distinguish between LFC and HFC calls. We assessed model performance based on their ability to classify vocalizations from the fused acoustic features, reporting metrics such as accuracy, precision, recall, and F1-score. Through this fusion of feature extraction and machine learning, we developed subtle measures to interpret cows' emotional states based on their vocal expressions.

3.4 Sentiment Analysis

Sentiment analysis was integrated to examine the emotional content in cow vocalizations further, fusing acoustic features with semantic information. Using Python's Natural Language Toolkit (NLTK), we performed sentiment analysis on the transcriptions provided by the OpenAI Whisper model. Vocalizations were coded with emotional polarity—positive or negative—based on predefined acoustic features reflecting specific affective states. High-frequency calls (HFC) were labeled as reflecting distress or arousal, while low-frequency calls (LFC) were labeled as reflecting contentment or calm.

This sentiment analysis complemented the acoustic feature-based classification, reinforcing the distinction between positive (calm/contentment) and negative (distress/arousal) affective states. By fusing semantic analysis with acoustic data, we enhanced the interpretive power of the vocalization data, aligning it with real-world behavioral cues to improve accuracy in assessing emotional states.

3.5 Cow Vocalization Ontology

We developed a cow vocalization ontology to categorize sounds based on specific fused acoustic features and relate them to corresponding emotional states. This ontology provides a structured framework for understanding how cows communicate distress or contentment through sound, offering a noninvasive means of assessing and monitoring cow welfare by analyzing frequency, loudness, and duration of HFC and LFC.

3.5.1 Acoustic Features of Cow Vocalizations

High-Frequency Calls (HFC): Associated with elevated emotional arousal, such as stress or discomfort, HFCs have frequencies ranging from 110.59 Hz to 494.16 Hz. Loudness levels range between –39.71 dB and –2.45 dB, making them louder compared to LFCs and likely to trigger urgency or discomfort. Durations range from 0.638 to 9.581 seconds, with longer calls indicating prolonged distress.

Low-Frequency Calls (LFC): Signifying contentment and calmness, LFCs have frequencies from 72.61 Hz to 183.27 Hz, producing lower and soothing tones during relaxed emotional states. Loudness levels range from –53.88 dB to –8.16 dB, reflecting a quieter communicative style compared to HFCs. Durations range from 0.650 to 2.921 seconds, representing brief interactions in relaxed situations.

3.5.2 Emotional Associations

By categorizing vocalizations as HFC and LFC through fused acoustic features, we infer the following emotional associations:

High-Frequency Calls: Linked to distress or arousal, these louder, longer, and higher-pitched sounds are associated with fear, isolation, or physical discomfort, serving as indicators of stress and used for long-distance communication.

Low-Frequency Calls: Correlating with contentment or calmness, these soft and low-pitched sounds are produced when cows are relaxed, particularly near their calves, used for close-contact communication.

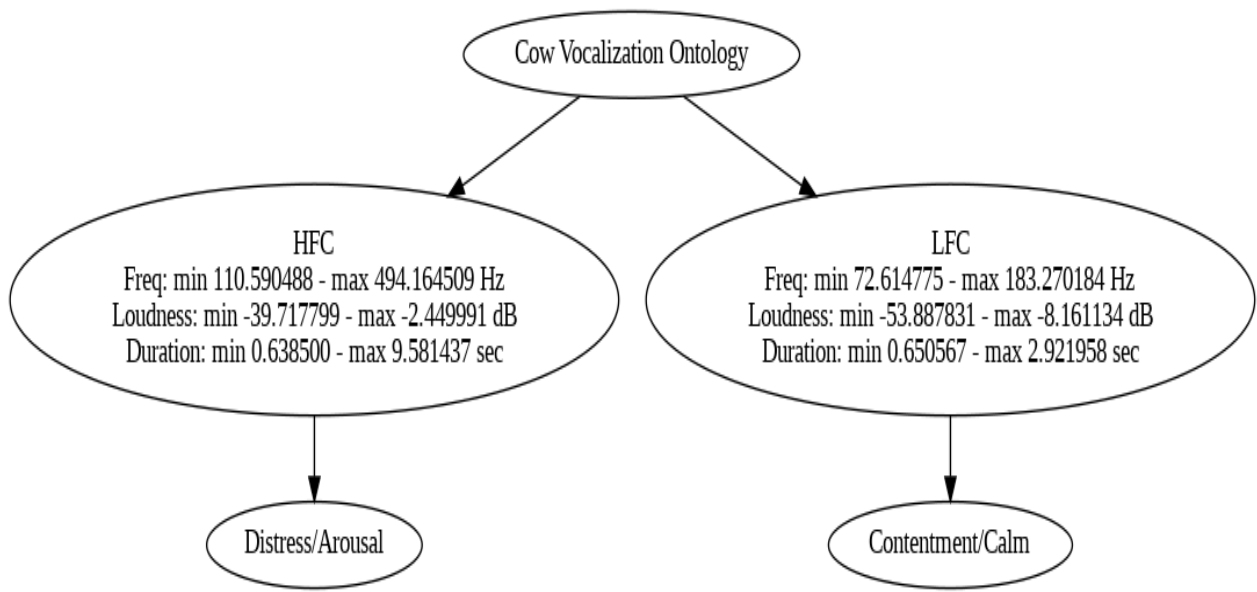

**Figure 1: Cow Vocalization Ontology - A Framework for Understanding Emotional States**

Figure 1 illustrates the ontology of cattle vocalizations, classifying them based on acoustic properties into HFCs and LFCs. HFCs are characterized by higher frequencies (110.59–494.16 Hz), louder loudness (–39.71 to –2.45 dB), and longer durations (0.638–9.581 seconds), typically associated with distress or arousal. In contrast, LFCs exhibit lower frequencies (72.61–183.27 Hz), quieter loudness (–53.88 to –8.16 dB), and shorter durations (0.650–2.921 seconds), reflecting contentment or calmness. This framework facilitates the classification of cow vocalizations into emotional contexts based on their acoustic features.

## 4. Results

*4.1 Bigram Analysis*

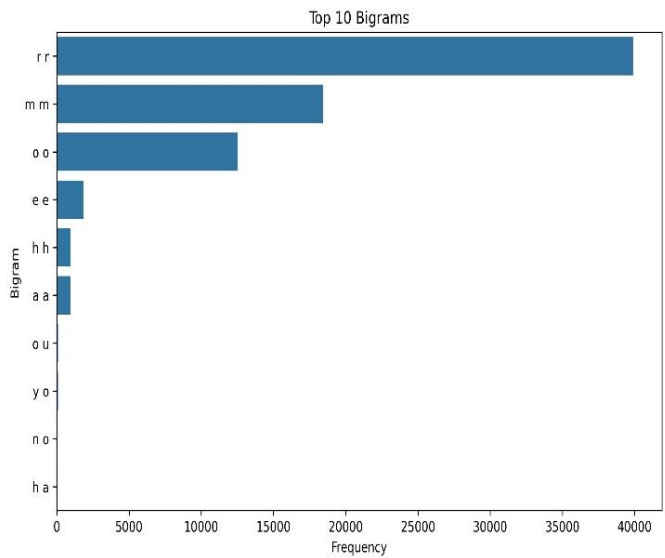
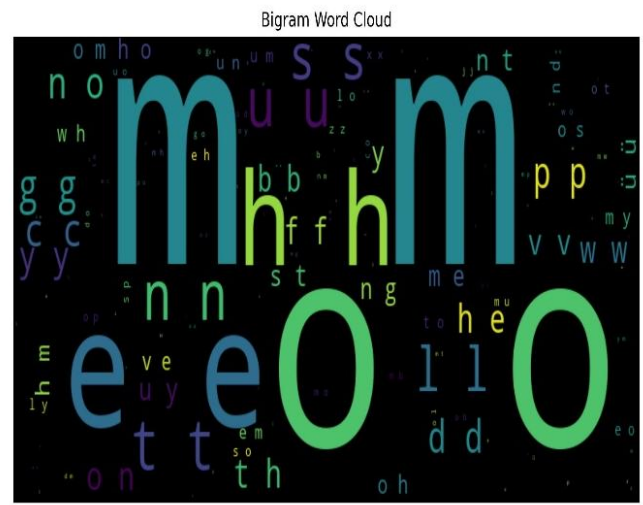

**Figure 2. Top 10 Most Frequent Bigram**              **Figure 3. Top 10 Most Frequent Bigram Word Cloud**

Figure 2 illustrates the bigram frequency analysis, while Figure 3 presents the corresponding word cloud. The bigram analysis effectively captures recurring patterns within the cow vocalization dataset, revealing significant insights when fused with acoustic data. Notably, the bigram "rr" is the most frequent, occurring approximately 40,000 times, closely followed by "mm" and "oo." These frequent bigrams suggest repetitive sound patterns inherent in cow vocalizations, potentially corresponding to specific emotional states or communication intents. The prominence of these bigrams indicates rhythmic or repetitive vocal elements that, when integrated with acoustic features, enhance our understanding of bovine communication dynamics. Recognizing these patterns through multi-source data fusion allows for more accurate interpretation of behavioral contexts and emotional states in cows.

*4.2 Unigram Word Cloud*

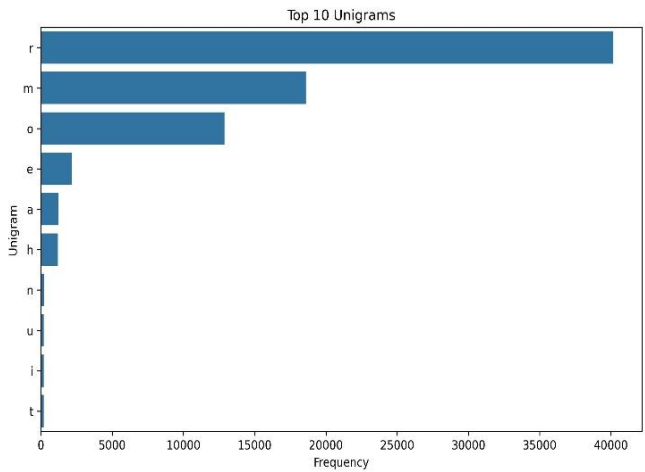
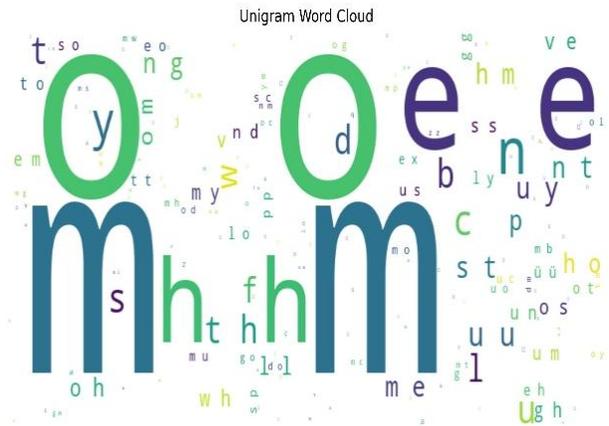

**Figure 4. Top 10 Most Frequent Unigram**

**Figure 5. Unigram Word Cloud**

Figures 5 and 6 display the unigram analysis and the unigram word cloud, respectively, providing critical information about character distribution within the dataset. The character "r" emerges as the most frequent, appearing nearly 40,000 times, followed closely by "m," which likely corresponds to specific vocalizations in bovine communication. High frequencies of characters like "o," "e," and "a" may represent common phonations in cow vocalizations. In contrast, less frequent characters such as "h," "n," "u," "i," and "t" suggest less common sounds or utterances. This significant variation in character frequency highlights the linguistic complexity and diversity within cow vocalizations. By fusing this linguistic data with acoustic features, we gain deeper insights into communication behaviors, facilitating the development of more sophisticated models for emotional state detection.

*4.3. Acoustic Feature Analysis*

Here we explore the acoustic properties of bovine vocalizations through multi-source data fusion, integrating spectral, temporal, amplitude and energy, formant, and prosodic analyses. By examining both high-frequency calls (HFC) and low-frequency calls (LFC), we aim to understand the relationship between acoustic patterns and emotional states such as distress, arousal, or contentment. Fusing these acoustic features provides an objective basis for classifying bovine vocal behaviors, enhancing our understanding of herd communication dynamics and supporting welfare assessments based on vocal indicators.

*4.3.1 Spectral Analysis*

Spectral analysis decomposes signals into their constituent frequencies using techniques like the Fourier Transform, identifying periodic patterns and noise characteristics (Lange et al., 2021). In cow vocalizations, spectral analysis is crucial for understanding how cows express emotions like pain, stress, or discomfort. By fusing frequency and acoustic properties, we can identify specific vocal structures associated with different behavioral states. For instance, stressful vocalizations often exhibit unique spectral features, such as increased harmonic content or variable pitch (Ghaderpour et al., 2021). This fusion of spectral data aids in developing monitoring systems to improve animal health and welfare (Vidana-Vila et al., 2023).

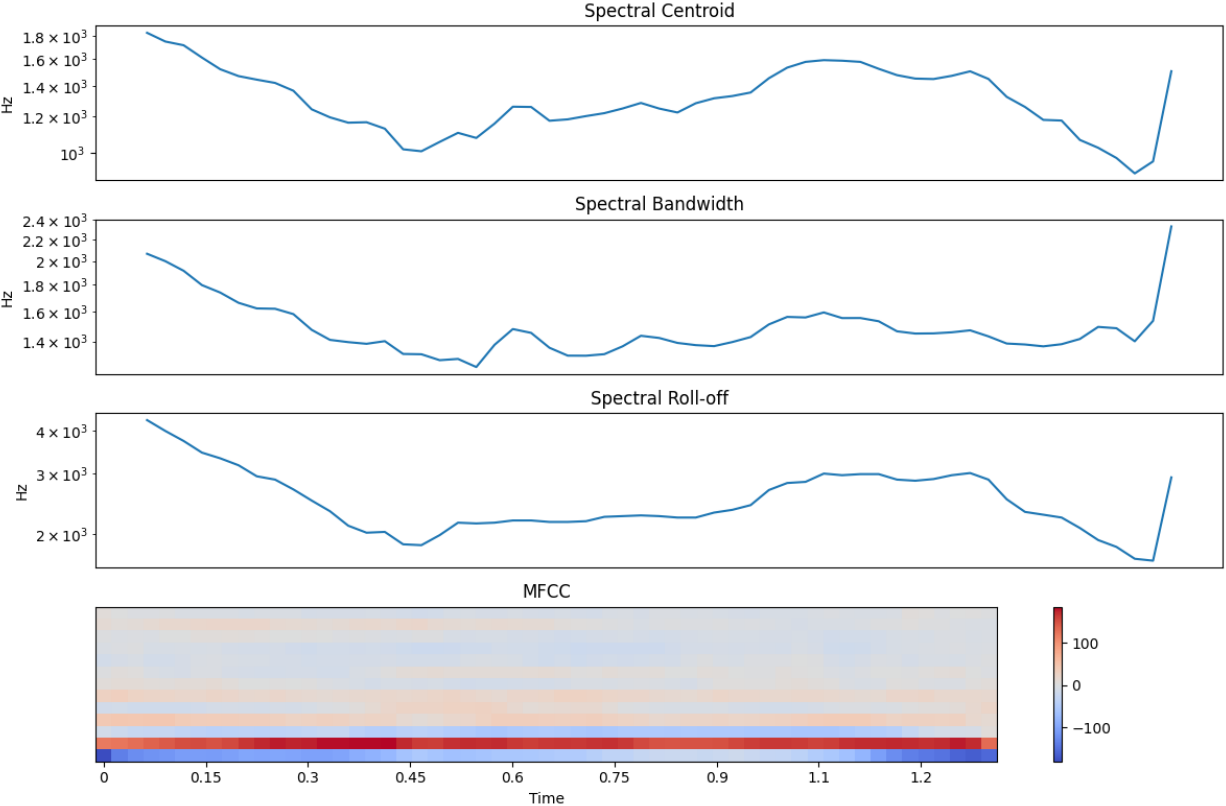

Figure 6: Spectral Analysis of a typical Low-Frequency Calls in Cows Vocalization

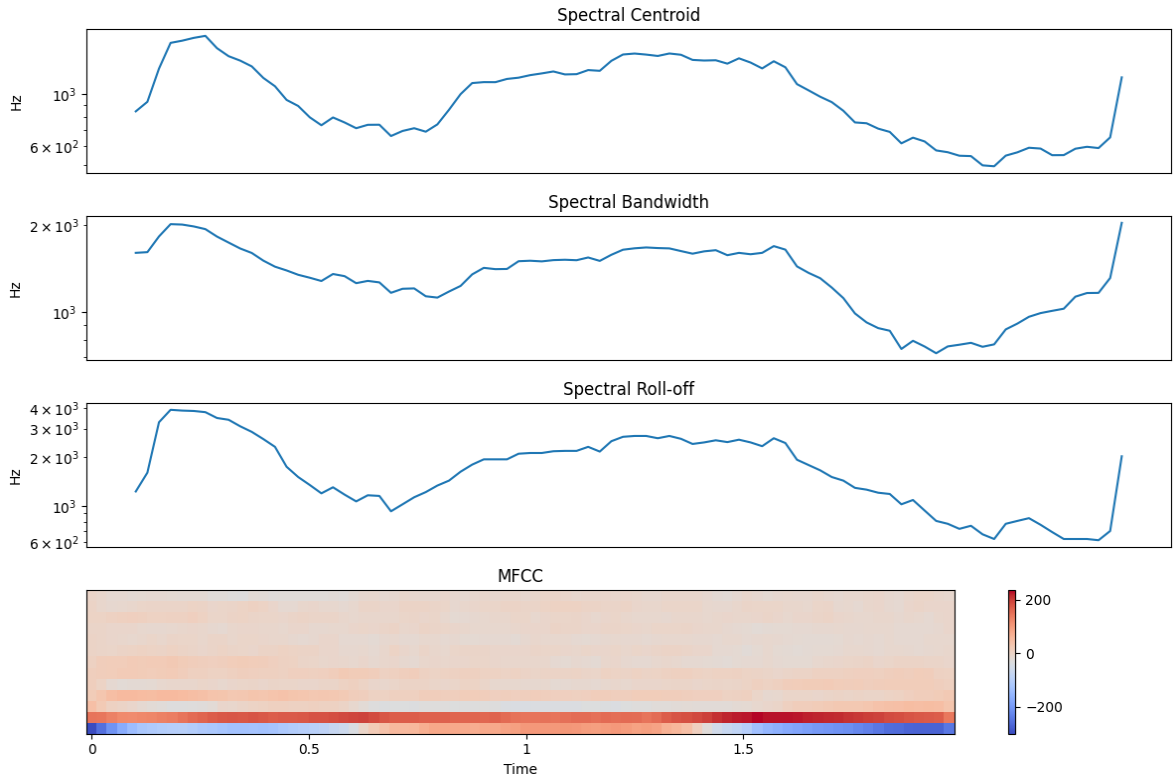

Figure 7: Spectral Analysis of a typical High-Frequency Calls in Cows Vocalization

Figure 6 shows that LFCs have lower spectral centroids (1,000–1,800 Hz) and narrower bandwidths, with energy concentrated below 3,000 Hz, indicating calm or routine communication. In contrast, Figure 7 reveals that HFCs possess higher spectral centroids (600–3,000 Hz) and wider bandwidths, with energy extending beyond 4,000 Hz, signifying urgent or distress-related vocalizations. Mel-frequency cepstral coefficients (MFCCs) for LFCs display smoother transitions, while HFCs exhibit abrupt changes, reflecting their complex and intense nature. By fusing spectral characteristics with behavioral data, we enhance the automatic detection of cow vocalizations for monitoring welfare and emotional states.

*4.3.2 Temporal Analysis*

Temporal analysis examines how vocal patterns vary over time, providing insights into cow behavior, health, and emotions. Parameters like duration, timing, and repetition are critical for understanding communication under different conditions (Hernández-Castellano et al., 2023). Temporal patterns, when fused with other acoustic features, enable the detection of behaviors related to feeding, stress, and herd dynamics (Gavojdian et al., 2024), enhancing computerized monitoring systems through information fusion.

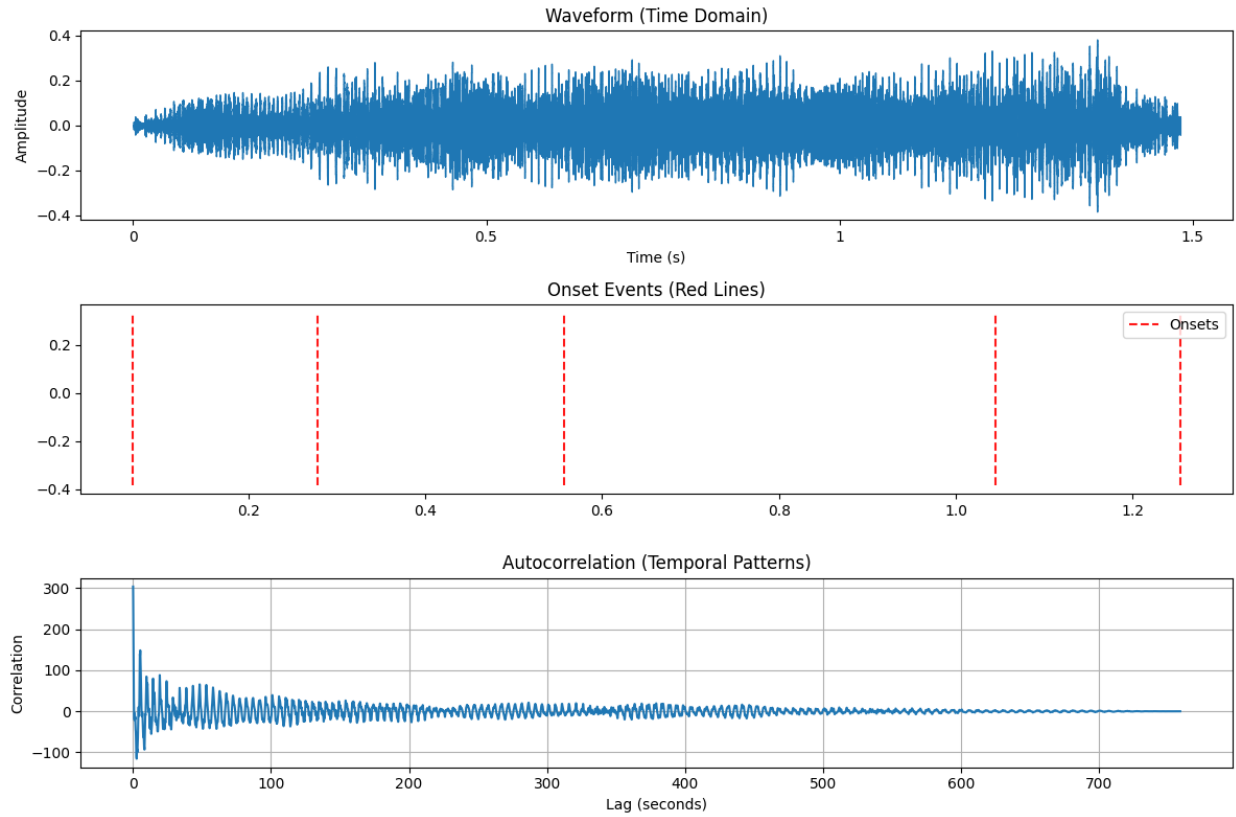

Figure 8: Temporal Analysis of a typical Low-Frequency Calls in Cows Vocalization

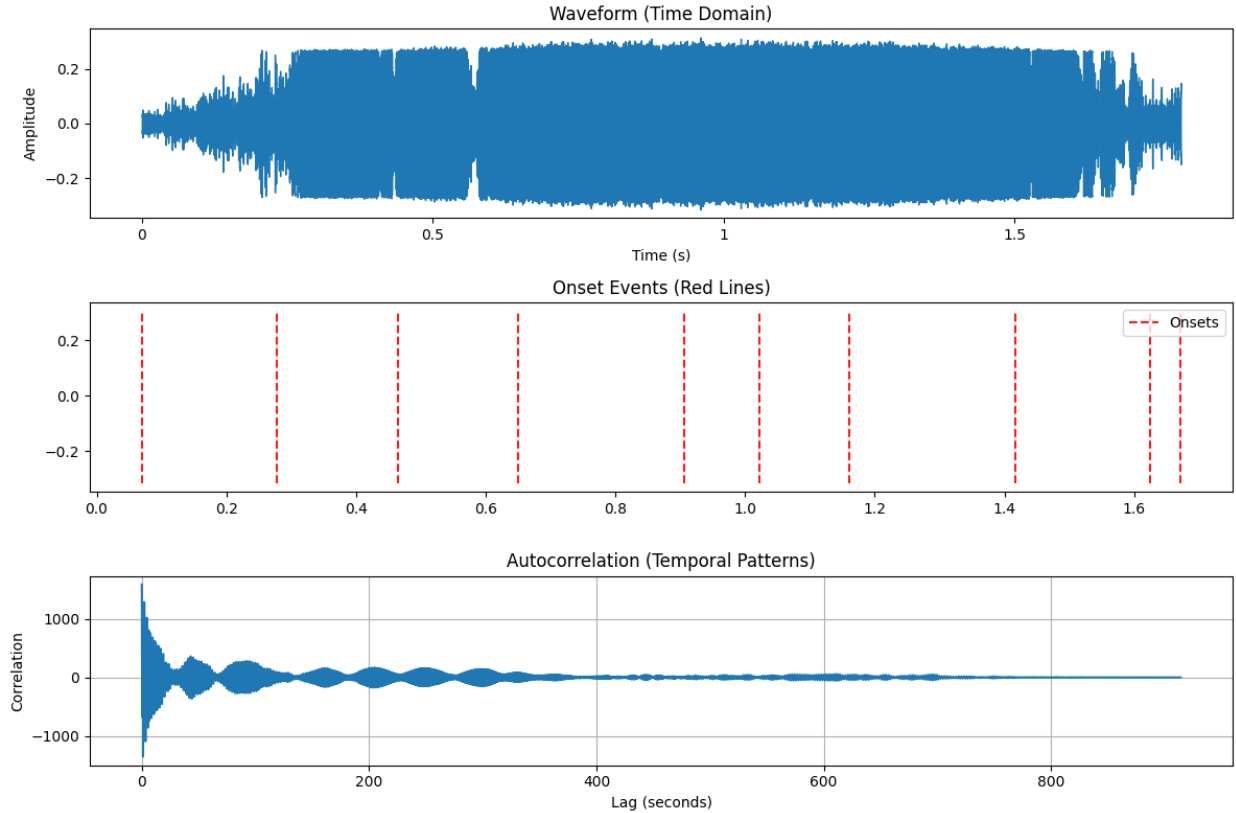

Figure 9: Temporal Analysis of typical High-Frequency Calls in Cows Vocalization

Figures 8 and 9 demonstrate significant differences in vocalization patterns. The HFC audio lasts 1.79 seconds with 10 detected events, indicating rapid, frequent vocalizations with an average interval of 0.18 seconds, suggesting urgency and potential stress. Conversely, the LFC audio spans 1.48 seconds with only 5 events and a mean interval of 0.30 seconds, indicating a more tranquil communication mode. Fluctuations in HFC intervals, with a minimum of 0.0464 seconds, are often linked to distress, while the stable rhythm of LFCs implies social connection and satisfaction. Fusing these temporal features with other acoustic data enhances our ability to monitor cattle welfare and understand their communication and emotional states.

*4.3.3 Amplitude and Energy Analysis*

Amplitude, measured in decibels, and energy quartiles (Q25%, Q50%, Q75%) provide information about the intensity and frequency distribution of vocalizations (Briefer, 2012). Variations in these parameters are associated with different emotional responses (Manteuffel et al., 2004) and, when fused with other acoustic features, offer valuable insights into animal welfare (Gavojdian et al., 2024). Figures 10 and 11 reveal distinct differences between LFC and HFC calls. The mean root mean square (RMS) energy for LFCs is 0.0934, indicating low intensity associated with calm behaviors. In contrast, HFCs have a mean RMS energy of 0.1887, reflecting more intense calls linked to stress. The zero-crossing rate averages further distinguish the calls; LFCs have a rate of 0.0427, indicating smoother transitions, while HFCs have a higher rate of 0.0492, suggesting complex emotional expressions. By fusing amplitude and energy data with other features, we can better assess emotional and behavioral contexts, essential for welfare evaluations.

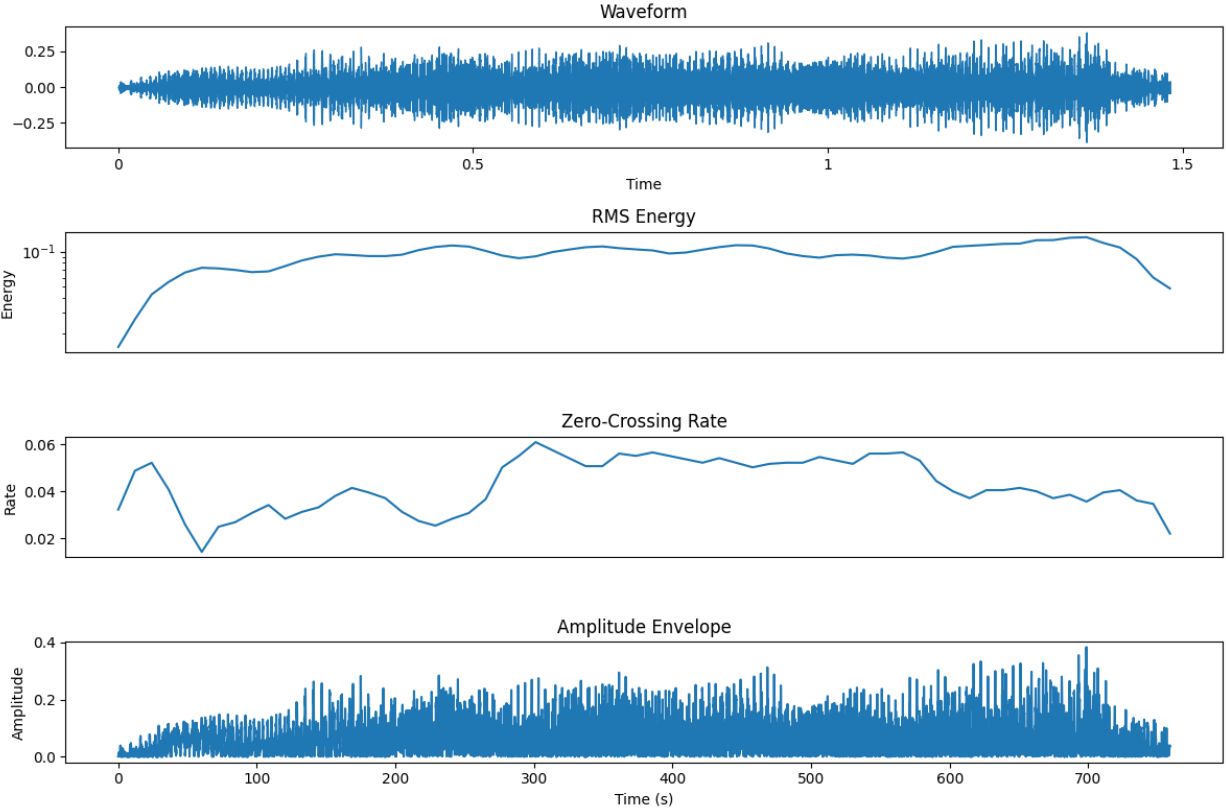

**Figure 10: Amplitude and Energy Analysis of typical Low-Frequency Calls in Cows Vocalization**

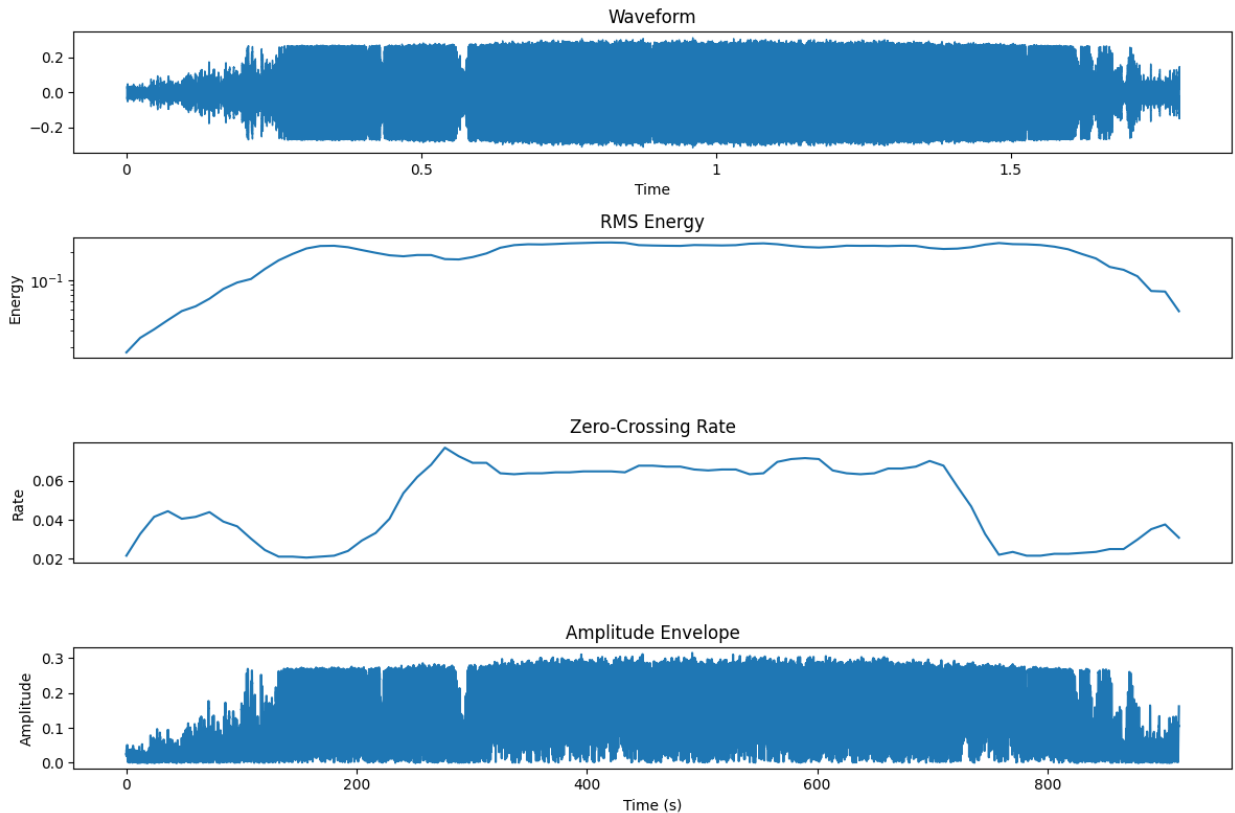

**Figure 11: Amplitude and Energy Analysis of typical High-Frequency Calls in Cows Vocalization**

*4.3.4 Formant Analysis*

Formants are frequencies corresponding to vocal tract resonances, critical for understanding vocalization characteristics (Morton, 1977). Key parameters include mean frequencies (F1mean, F2mean, F3mean, F4mean) and their ranges, indicating the frequency distribution of each formant (Moshou et al., 2001). Formant analysis, when fused with other acoustic features, aids in identifying vocalization types and emotional states (Nordell & Valone, 2017). Figures 12 and 13 show that F1 frequencies for HFC (609.56 Hz) and LFC (617.35 Hz) are similar, suggesting a consistent vocal tract configuration. However, F2 and F3 differ; HFCs have a higher F2 frequency (1,704.81 Hz) compared to LFCs (1,542.96 Hz), possibly indicating vocal tract constriction associated with distress. LFCs have a slightly higher F3 (2,844.92 Hz vs. 2,779.11 Hz), reflecting stable, relaxed communications. Fusing formant data with other acoustic features enhances our understanding of emotional contexts, aiding in welfare assessments.

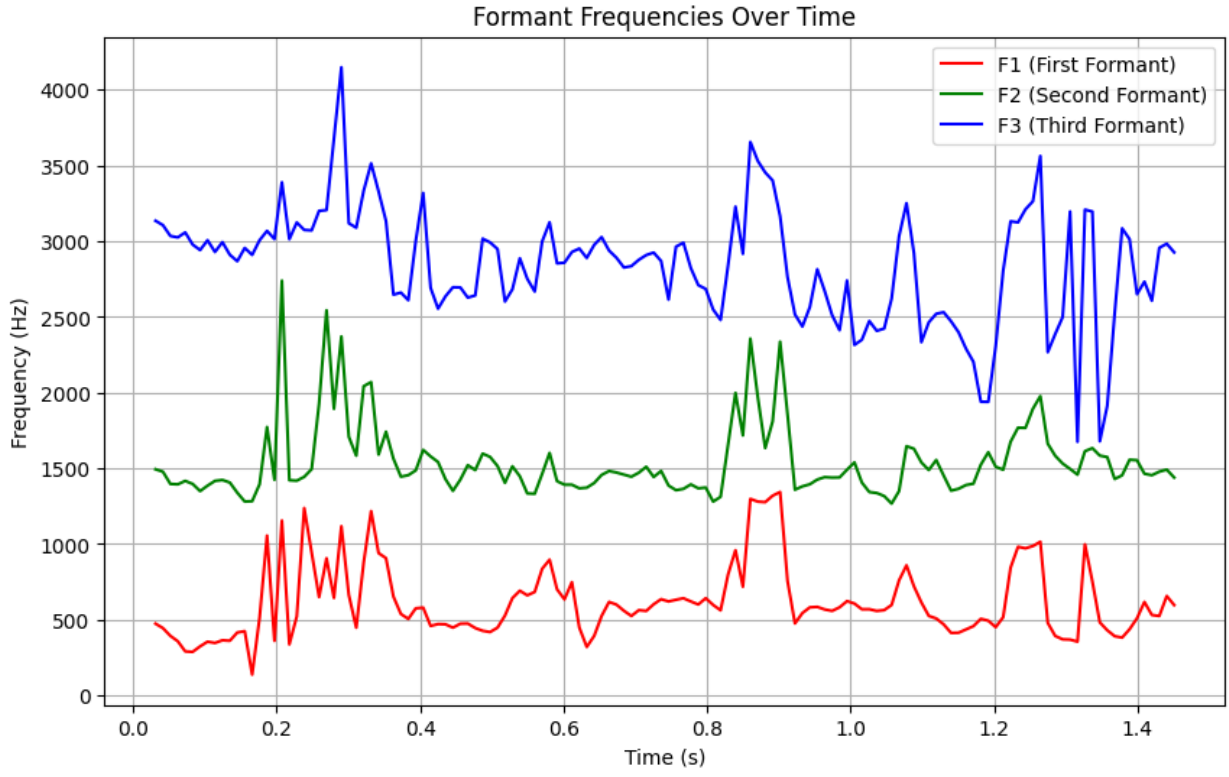

**Figure 12: Formant Analysis of typical Low-Frequency Calls in Cows Vocalization**

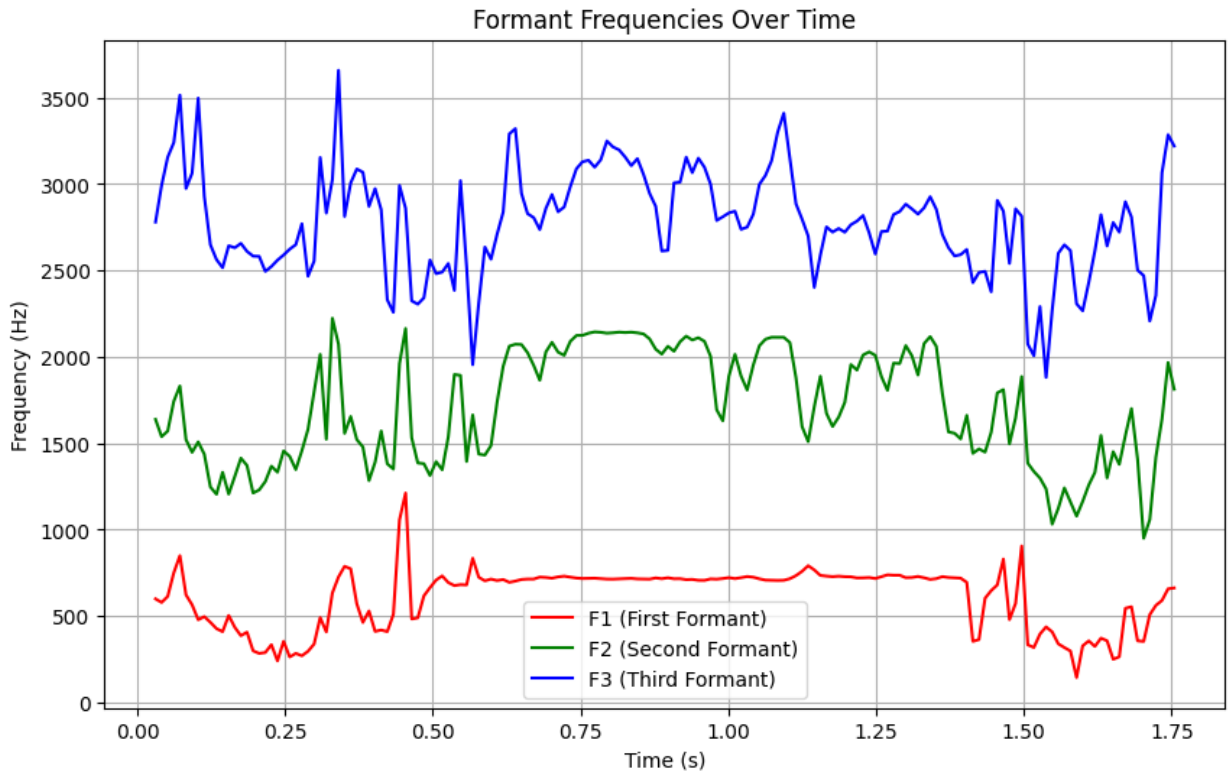

**Figure 13: Formant Analysis of typical High-Frequency Calls in Cows Vocalization**

*4.3.5 Prosodic Analysis*

Prosodic analysis examines features like pitch, loudness, tempo, and rhythm, contributing to the emotional and communicative aspects of vocalizations. Variations in prosodic features can indicate emotional states, crucial for understanding animal welfare (Friel et al., 2019; Meen et al., 2015). The prosodic analysis reveals clear differences between HFC and LFC calls. HFCs exhibit a broad amplitude range of 514.20 Hz, with high variability in fundamental frequency (F0), reflecting urgency, stress, or excitement. LFCs display a limited pitch range of 33.03 Hz, indicating stable vocal expressions associated with calmness and social bonding. By fusing prosodic features with other acoustic data, we enhance our ability to recognize vocal patterns, understand cattle behavior, and improve welfare through informed management practices.

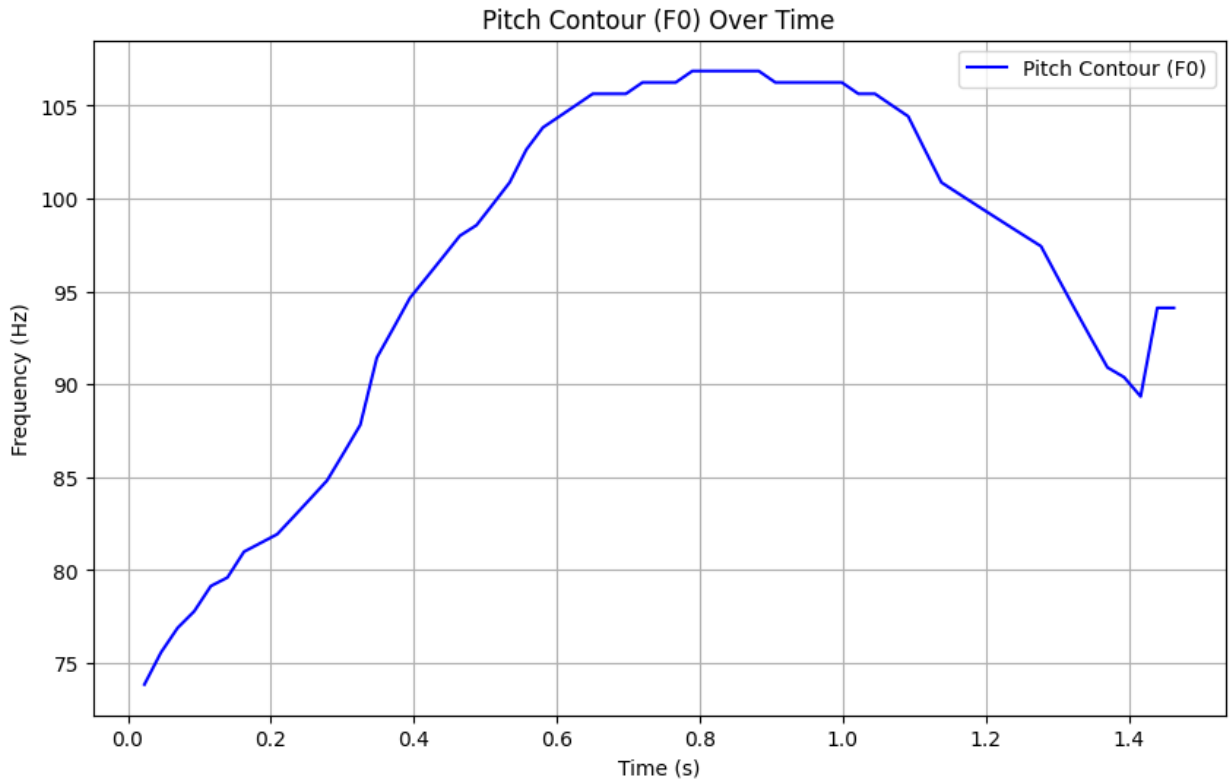

**Figure 14: Prosodic Analysis of typical Low-Frequency Calls in Cows Vocalization**

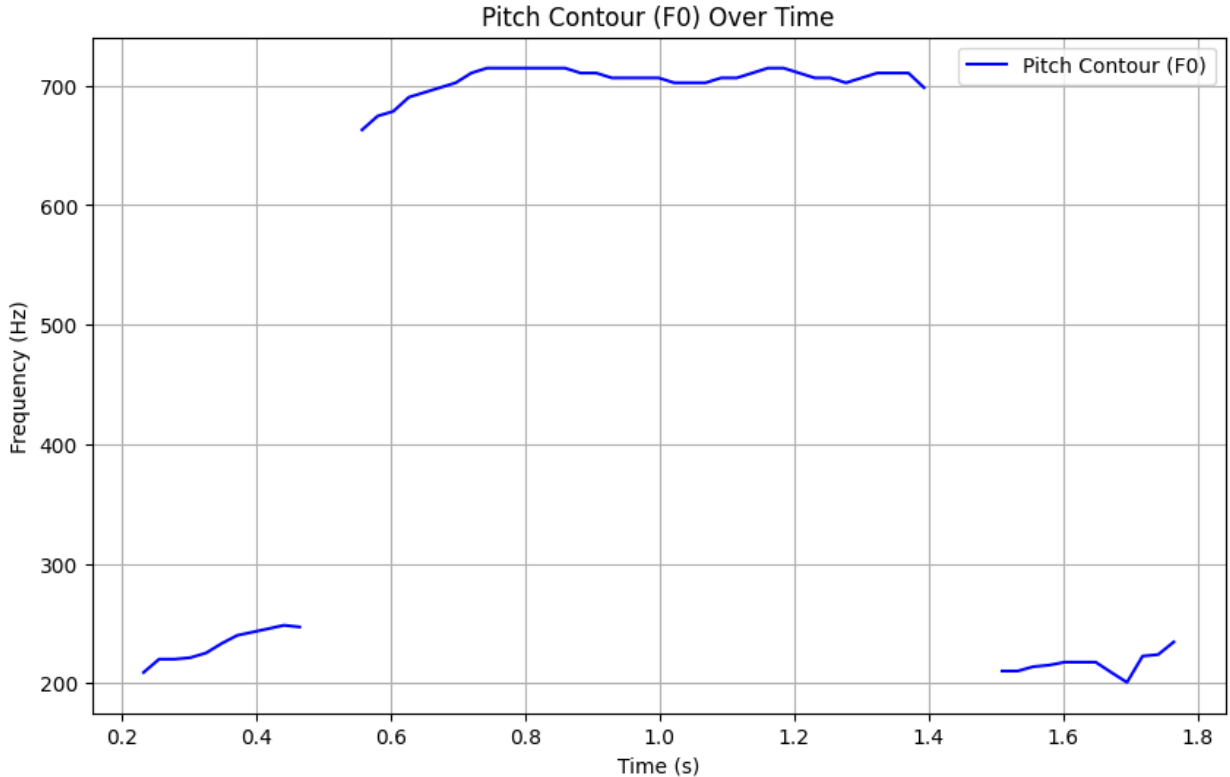

Figure 15: Prosodic Analysis of typical High-Frequency Calls in Cows Vocalization

## 5. Fusion-Based Machine Learning Models: Architectures and Performance Evaluation

### 5.1 Random Forest

The Random Forest model, an ensemble learning method, builds multiple decision trees on random subsets of data and aggregates predictions, making it robust against overfitting and noise. By fusing features like frequency, loudness, and duration, the model effectively handles diverse data. The Random Forest classifier correctly predicts 135 instances of Distress/Arousal with just one misclassification and accurately identifies 42 instances of Contentment/Calm, misclassifying two. With an accuracy of 97.25% and an AUC of 0.99, the model (Figure 16 and 17) demonstrates strong performance by effectively fusing acoustic features for classification.

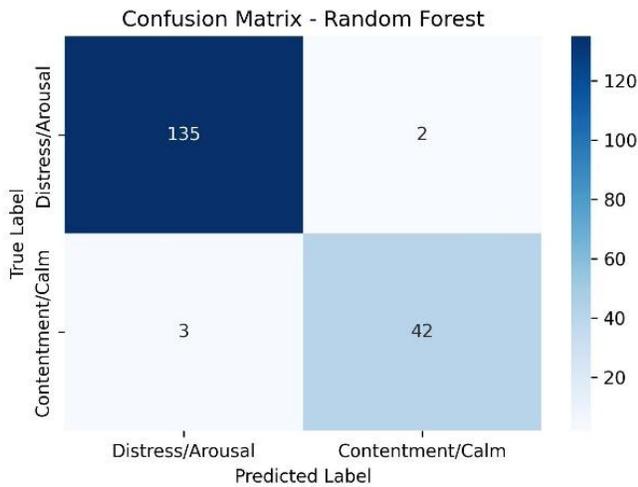

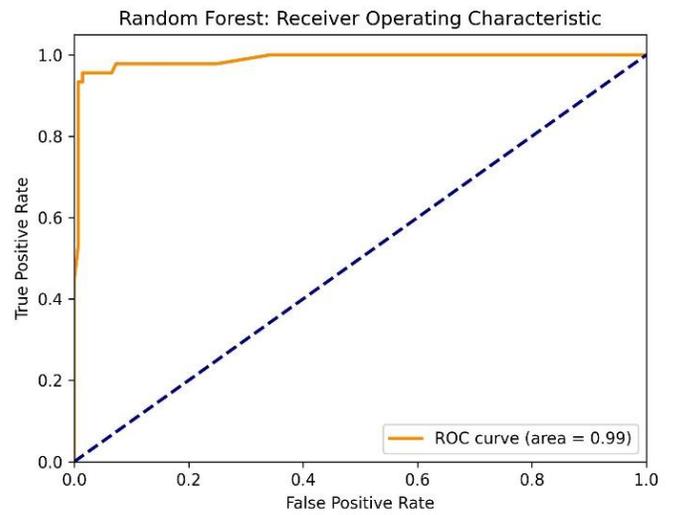

Figure 16 Confusion Matrix of Random Forest    Figure 17 Receiver Operating Curve of Random Forest

*5.2 Support Vector Machine*

The SVM model employs a linear kernel to classify vocalizations using fused features. It finds the optimal hyperplane that separates classes, maximizing the margin between them. The SVM classifier successfully differentiates between emotional states, correctly predicting 136 instances of Distress/Arousal (Figure 18 and 19) and 43 instances of Contentment/Calm, achieving an accuracy of 98.35%. With an AUC of 0.99, the model's performance highlights the effectiveness of fusing acoustic features for accurate classification.

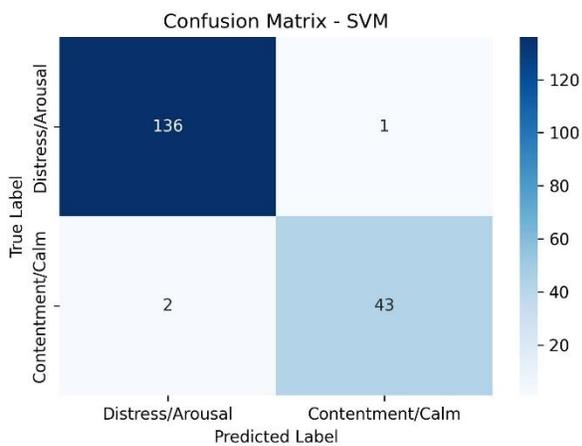

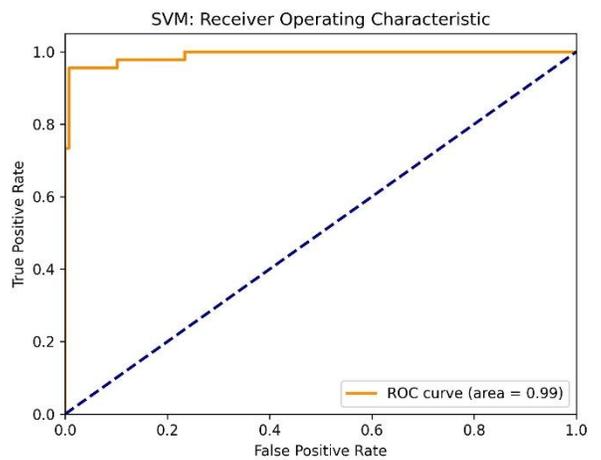

Figure 18 Confusion Matrix of SVM    Figure 19 ROC Curve of the SVM Model

## 5.3 Recurrent Neural Network

The RNN model captures sequential relationships inherent in vocalization data. Its architecture includes a SimpleRNN layer with 32 units and a Dense output layer, processing temporal evolutions of fused acoustic features. The RNN model performs well in identifying Distress/Arousal calls but struggles with Contentment/Calm, possibly due to dataset size and sequence length limitations. With an accuracy of 88% and an AUC of 0.96 for both classes, the model's performance (Figure 20 and 21) indicates the need for further refinement, possibly through enhanced data fusion techniques.

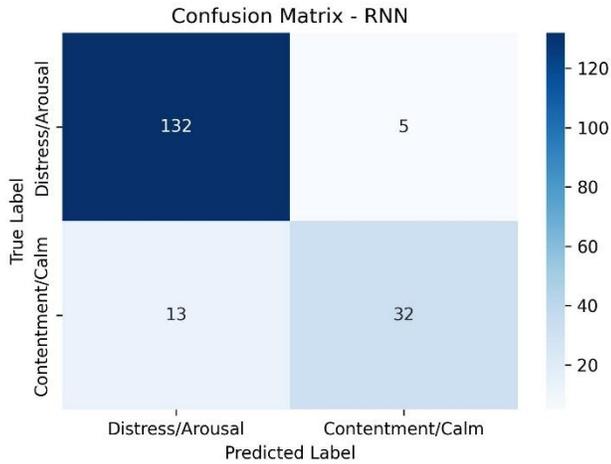

**Figure 20 Confusion Matrix of RNN Model**

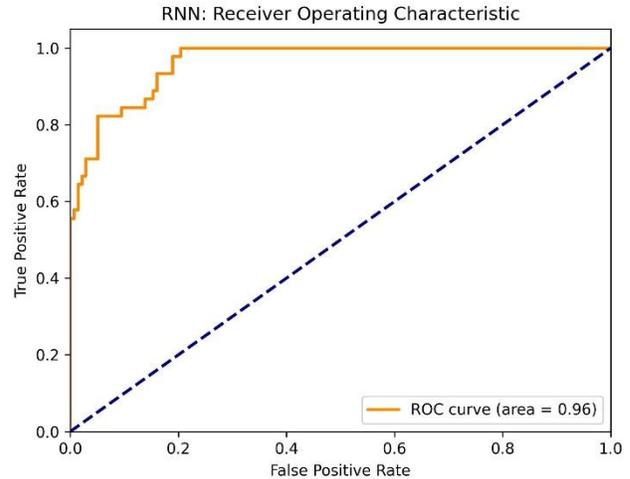

**Figure 21 ROC Curve of the RNN Model**

## 5.4 Classification Performance

Table 2 presents the precision, recall, F1-score, and accuracy for each model across the two emotional categories. The SVM model outperforms others, showing a well-balanced performance across both emotional classes, effectively utilizing fused acoustic features. The Random Forest model also performs well but shows slight difficulty with Contentment/Calm classification. The RNN model's performance suggests that incorporating more advanced fusion techniques and larger datasets could enhance its effectiveness.

Table 2: Performance Evaluation Result of Random Forest, Support Vector Machine and RNN

| Model | Class | Precision | Recall | F1-Score | Accuracy |
|---|---|---|---|---|---|
| **Random Forest** | Distress/Arousal | 0.98 | 0.99 | 0.98 | **0.9725** |
| | Contentment/Calm | 0.95 | 0.93 | 0.94 | |
| | Macro Avg | 0.97 | 0.96 | 0.96 | |
| | Weighted Avg | 0.97 | 0.97 | 0.97 | |
| **SVM** | Distress/Arousal | 0.99 | 0.99 | 0.99 | **0.9835** |
| | Contentment/Calm | 0.98 | 0.96 | 0.97 | |
| | Macro Avg | 0.98 | 0.97 | 0.98 | |
| | Weighted Avg | 0.98 | 0.98 | 0.98 | |
| **RNN** | Distress/Arousal | 0.89 | 0.97 | 0.93 | **0.88** |
| | Contentment/Calm | 0.88 | 0.62 | 0.73 | |
| | Macro Avg | 0.88 | 0.80 | 0.83 | |
| | Weighted Avg | 0.88 | 0.88 | 0.88 | |

*5.5  Feature Importance*

Feature importance enhances model interpretability and aids in feature selection, critical in machine learning (Liaw & Wiener, 2002; Guyon & Elisseeff, 2003). Understanding which variables contribute most to predictions allows for better insights into model behavior and supports the development of informed strategies (Chen & Guestrin, 2016; Molnar, 2019). The Random Forest model identifies frequency as the most important predictor (score of 0.70), indicating its critical role in distinguishing vocalizations. Loudness follows with a score of 0.22, while duration ranks lower at 0.09 (Figure 22) but still contributes to classification. By understanding feature importance through data fusion, we can target interventions, inform breeding decisions, and develop better policies for animal welfare, ultimately improving outcomes for cows and the agricultural industry.

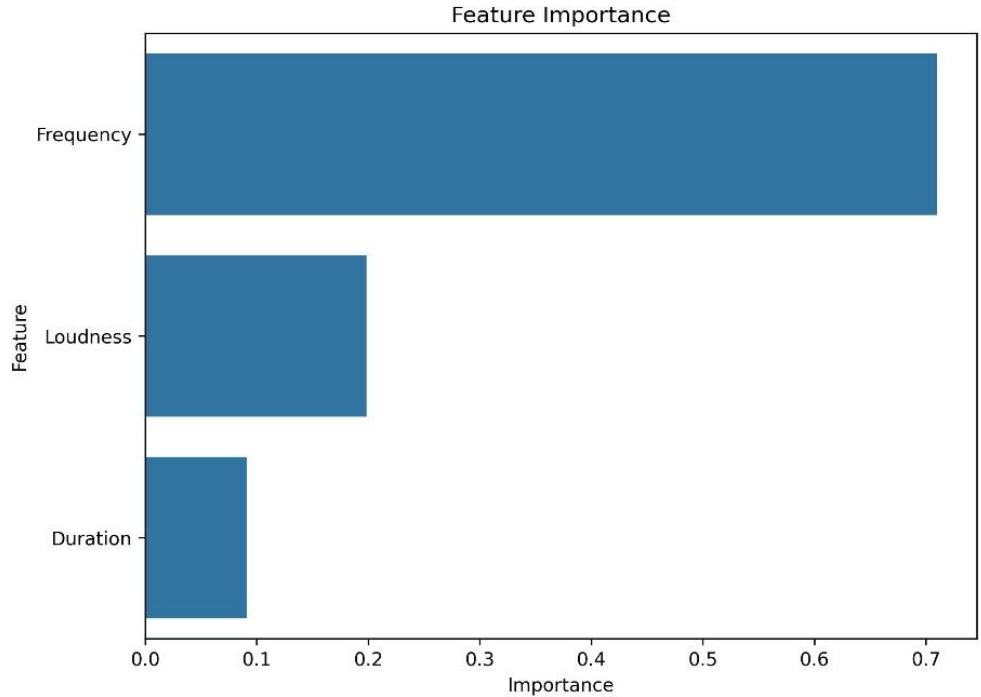

Figure 21 Feature Importance of the Random Forest Model

## 6. Discussions

This study introduces a novel approach to animal behavior and welfare assessment by employing multi-modal information fusion techniques to analyze cow vocalizations. By integrating acoustic feature extraction with advanced machine learning algorithms, we classified cow calls into two primary categories: high-frequency calls (HFCs), associated with stress or arousal, and low-frequency calls (LFCs), indicative of contentment or calmness. While this binary classification provides a foundational framework for interpreting emotional states, it is crucial to acknowledge that bovine emotional experiences are inherently complex and multifaceted.

By fusing key acoustic features—frequency, loudness, and duration—we utilized Random Forest, Support Vector Machine (SVM), and Recurrent Neural Network (RNN) models to classify cow vocalizations effectively. The SVM model excelled, achieving an accuracy of 98.35%, while the Random Forest model demonstrated robust performance with an F1-score of 0.98, particularly in predicting distress-related calls. The RNN model faced challenges in classifying calm vocalizations, potentially due to limitations in capturing temporal dependencies and the relatively small dataset. This highlights the importance of data fusion techniques in enhancing model performance by integrating multiple acoustic features.

Feature importance analysis underscored frequency as the most critical predictor, followed by loudness and duration, emphasizing the integral role of these fused acoustic properties in distinguishing cows' emotional states. Detailed acoustic analyses provided deeper insights: spectral analysis revealed that HFCs possessed higher centroids and bandwidths, correlating with urgent or distress calls, while LFCs exhibited smoother transitions and energy distributions, associated with calm or social interactions. Temporal analysis indicated that HFCs were characterized by shorter intervals and rapid vocalization sequences, reflecting heightened arousal, whereas LFCs displayed longer, stable intervals indicative of relaxed states.

Amplitude and energy analyses reinforced these findings; HFCs exhibited higher root mean square (RMS) energies and more variable zero-crossing rates, signifying the intensity of distress calls. Conversely, LFCs demonstrated lower amplitude and energy, aligning with contentment or routine communications. Formant analysis further revealed that HFCs presented higher second formants, suggesting vocal tract constriction during distress, whereas LFCs showed higher third formants, associated with stable, relaxed communications. By fusing these acoustic features, we enhanced the accuracy of emotional state classification, showcasing the power of information fusion in bioacoustic analysis.

Integrating sentiment analysis into the Whisper model added an extra layer of classification by correlating transcribed vocalizations with emotional polarity through multi-source data fusion. This non-invasive monitoring approach holds significant potential in precision livestock farming, enabling early detection of stress or discomfort in cows. The fusion of bioacoustic analysis and advanced machine learning algorithms demonstrates great promise for developing automated animal behavior monitoring systems. Our framework leverages fused acoustic and semantic features for real-time welfare evaluation, offering a more objective, evidence-based approach to enhancing animal care, productivity, and management practices.

## 7. Future Directions

Building on our findings, this study opens significant avenues for future research and practical applications in animal welfare and livestock management. One promising direction is to enrich the dataset by incorporating diverse cow breeds and crossbreeds from various environments, enhancing the generalization capabilities of the machine learning models. By fusing data from multiple sources and conditions, we can address limitations related to breed-specific vocalizations and environmental influences, further refining our models through comprehensive information fusion.

Applying advanced deep learning frameworks and transformer models in bioacoustic studies could further enhance the robustness and accuracy of vocalization classification. These models are capable of capturing complex patterns and temporal dependencies in vocal data, leading to more precise emotional state assessments. Developing a comprehensive cow welfare evaluation system that integrates multiple modalities—beyond vocalizations—is another promising direction. By fusing visual and behavioral data, such as posture, movement patterns, and facial expressions, with acoustic signals, we can achieve a more holistic and reliable estimate of animal welfare. This multimodal data fusion would significantly enhance the sensitivity and specificity of welfare assessments.

Implementing these models in real-time monitoring systems on farms presents a compelling opportunity. By integrating and fusing data from multiple sensors and sources in real-time, automated recognition of stress or discomfort signs becomes feasible, enabling immediate interventions. Such systems could revolutionize livestock management by facilitating proactive care and early detection of welfare issues, ultimately improving animal welfare and farm productivity. The fusion of real-time data streams aligns with the journal's emphasis on multi-sensor information fusion and its applications.

Further research could explore the emotional content of other call types, including intermediate frequencies and complex vocal patterns, to unravel the dynamics of bovine communication. By fusing a broader spectrum of vocalizations with other data types, we can deepen our understanding of cows' emotional expressions and social interactions. Incorporating cross-validation with independent physiological stress indicators, such as heart rate or cortisol levels, would strengthen the interpretation of vocalization-derived emotional states. This multimodal data fusion and validation approach would enhance the reliability and robustness of welfare assessments.

We recommend that future studies develop more sophisticated systems capable of classifying a wide range of emotions and contexts in cow vocalizations through advanced information fusion techniques. By leveraging multi-sensor, multi-source data fusion, we can significantly advance our knowledge of bovine communication. Such advancements will not only contribute to improving animal welfare but also revolutionize monitoring practices in livestock management, leading to more ethical and efficient farming practices. Embracing the principles of information fusion, future research can address real-world challenges in agricultural systems, aligning with the journal's scope and contributing to the broader field of multi-sensor data integration and analysis.

## 8. Conclusions

This study demonstrates the efficacy of combining advanced machine learning techniques with multi-modal information fusion for monitoring cow vocalizations and assessing their emotional states. By classifying calls into high-frequency distress calls and low-frequency contentment calls based on fused acoustic features—frequency, loudness, and duration—we highlight the transformative potential of automated systems in enhancing animal welfare within precision livestock farming. The high accuracies achieved, particularly by the SVM and Random Forest models, underscore the urgent need to integrate acoustic and semantic analysis into routine farming practices. Our work paves the way for developing data-driven frameworks for animal care, leveraging information fusion to improve decision-making processes. The utilization of the Whisper model for cow vocalization transcription allowed for precise and efficient extraction of call data, facilitating a comprehensive fusion of acoustic and linguistic information. This transcription capability provides a foundation for more sophisticated processing and detailed analysis of vocalization patterns. While the Whisper model focuses on transcription, its integration within our fusion framework advances the goal of non-invasive, automated welfare monitoring through accurate classification of cow vocalizations. Future research will incorporate visual data, behavioral observations, and physiological measurements, enhancing the reliability of these systems through multimodal data fusion and advancing us toward fully automated precision farming.

Implementing such fusion-based systems will revolutionize how farmers detect signs of stress, discomfort, or illness in their herds, offering opportunities for early intervention to improve animal welfare and farm productivity. Integrating real-time acoustic monitoring empowers farmers to make informed and proactive decisions regarding their animals' health and well-being. As machine learning models evolve with larger and more diverse datasets, these tools will become capable of detecting even subtle emotional states and behaviors—a development of paramount importance in animal care. This study marks a significant step in utilizing bioacoustics and artificial intelligence for real-time welfare assessment, recognizing the crucial role of data-driven, fusion-based automated systems in advancing both animal welfare and productivity. As ethical and sustainable agricultural practices become increasingly essential, new technological developments in information fusion offer promising solutions for enhancing the quality of life for livestock without compromising their well-being, while also promoting efficient and responsible farming practices. Embracing multi-sensor and multi-source data fusion techniques not only improves the accuracy of welfare assessments but also aligns with the broader goals of sustainable and ethical farming. Our findings advocate for the integration of fusion-based technologies in livestock management, heralding a new era of intelligent, responsive, and humane agricultural systems.

**Author Contributions:** MM and GD conducted the farm trial and were responsible for the collection and analysis of the bioacoustics data. BJ analyzed and interpreted the results and wrote the initial draft of the manuscript. SN provided supervision, conceptualization, funding acquisition, coordinated the writing process, and performed manuscript revisions. All authors contributed critically to the writing of the manuscript and gave final approval for publication.

Acknowledgments: The vocalizations analyzed in this study were acquired during project number PN-III-P1-1.1-TE-2021-0027, supported by the Romanian Ministry of Research, Innovation and Digitization, CNCS—UEFISCDI, within PNCDI III. This work was also generously funded by the Natural Sciences and Engineering Research Council of Canada (RGPIN-2024-04450), Mitacs Canada (IT36514), the Department of New Brunswick Agriculture (NB2425-0025), and the Nova Scotia Department of Agriculture (NS-54163).